\documentclass[eqsecnum,aps,epsfig]{revtex4}
\usepackage{amsmath,amssymb,amsthm,bm}
\usepackage{psfrag}
\usepackage[dvips]{graphicx}

\begin{document}
\title{Short distance potential and the thick center vortex model}
\author{S.~Deldar$^\dagger$ and S.~Rafibakhsh$^\ddagger$}
\affiliation{
Department of Physics, University of Tehran, P.O. Box 14395/547, Tehran 1439955961,
Iran \\
$^\dagger$E-mail: sdeldar@ut.ac.ir\\
 $^\ddagger$E-mail: rafibakhsh@ut.ac.ir
}
\date{\today}

\begin{abstract}

The short distance potentials between heavy SU(3) and SU(4) sources are calculated by increasing the role of vortex fluxes piercing Wilson loops with contributions close to the trivial center element and by fluctuating the vortex core size in the model of thick center vortices. By this method, a Coulombic potential consistent with Casimir scaling is obtained. In addition, all other features of the potential including a linear intermediate potential in agreement with Casimir scaling and a large distance potential proportional to the $N$-ality of the representation are restored. Therefore, the model of thick center vortices may be used as a phenomenological model, which is able to describe the potential for all regimes.
\end{abstract}

\maketitle

\section{INTRODUCTION}\label{Intro}

The strong force has been known for a long time as one of the four fundamental forces of nature. However, many of its features are still unknown. Quarks are the fundamental and gluons are the mediating particles of the strong force. But, still no one knows the mechanism, which does not let them exist freely or in a color non-singlet state. Today, no one doubts that QCD confines quarks. When we try to take two quarks apart, the potential between them increases linearly by distance. This is the behavior of the potentials at intermediate and large distances, where the coupling constant is large, such that the perturbative methods can not be used in this regime. Besides lattice gauge theory which has been successful in producing the linear potential and therefore has confirmed confinement, there are some phenomenological models that try to explain the confinement mechanism. According to these models, the QCD vacuum is filled by some special class of field configurations. One type 
 of the candidates of these configurations are center vortices. These are line-like (in three dimensions) or surface-like (in four dimensions) objects, which carry a magnetic flux quantized in terms of the center elements of the gauge group. The vortex theory states that the area law for the Wilson loop is due to the quantum fluctuation in the number of center vortices linking the loop. The thick center vortex model has been fairly successful to reproduce the potentials of heavy quarks and higher representation sources at intermediate and large distances. Even though the perturbative method  gives the Coulombic contribution as a result of one gluon exchange at short distance, it would be very exciting if one could describe the potentials for all regimes with the same mechanism.

The model of thick center vortices has been originally introduced \cite{Faber:1997} as a simple model to calculate potentials between static sources. It gives a linear potential for intermediate distances and a correct asymptotic behavior for large distances. At intermediate distances, the string tensions for pairs of color sources in the group representation r, are proportional to the eigenvalues of the quadratic Casimir operator of the representation, which is in agreement with the lattice calculations. At large distances, the string tensions of zero $N$-ality representations are zero and the representations with the same non-zero $N$-ality acquire the same string tension. In this paper, we argue that, by modifying this model, one may even get a Coulombic behavior at short distances. The strength of the Coulombic part of the potential is proportional to the eigenvalue of the quadratic Casimir operator of the corresponding representation, as expected. We must mention that independent of this work, M. Faber and D. Neudecker are also working on the model to achieve a Coulombic potential . The preliminary results of that study are presented in reference \cite{faber2009}.

In the next section, we review briefly both thin and thick vortices and discuss the characteristics of the potentials obtained by the thick center vortex model. In section three, we discuss the role of the trivial and non-trivial center elements on the potentials. In section four, we  achieve the Coulombic part of the potential by increasing the role of contributions close to the trivial center element and show that it is proportional to the Casimir scaling.

\section{Vortices and confinement}

In the original vortex model \cite{Faber:1997}, it has been shown that the non-trivial center elements of the gauge group are responsible for the linear behavior of the potentials at large distances. By removing the non-trivial center elements of the gauge group from the lattice configurations, one would lose the linearity of the potentials \cite{Faber:1999}. $W(C)$ is the  Wilson loop in the SU($N$) gauge theory, with the contour C, which encloses the minimal area. We may consider it as being constructed of two main parts: one which is responsible for the linear potentials for both  intermediate and large distances, and another part which gives the short distance behavior,
\begin{equation}
<W(C)>= \exp[-\sigma(C)A]<W_{0}(C)>.
\label{W}
\end{equation}
Where $<W_{0}(C)>$ indicates the Coulombic contribution and the exponential part leads to a linear  potential. The center vortex model explains the linear behavior of the potential at intermediate and large distances by using the non-trivial center elements. Therefore, according to this model, the trivial center element may not contribute to the linear part of the potential. But, we know from the lattice gauge theory that the whole gauge group gives the complete potential. So increasing the role of the trivial center element may lead to the proper short distance behavior.

A Wilson loop is a gauge invariant Green function that describes the propagation of a static quark-antiquark pair, which is created within a distance $R$ at $t=0$ and propagates for the time $T$. One may find the heavy quark potential $V(R)$ by the Wilson loop at large $T$ by
\begin{equation}
W(R,t)\propto \exp^{-V(R)T}.
\label{WV}
\end{equation}
When the Wilson loop interacts with the vortex or in other words, when the vortex influences the worldline of the quark-antiquark potential, the average Wilson loop obeys the area law fall off rule for large enough time. This behavior implies a linear potential.

When the vortex is linked to a Wilson loop, the minimal surface is pierced by a center vortex.In this case, one may insert a non-trivial center element $\exp({\frac{2\pi i n}{N}})$ of the SU($N$) gauge group for a thin vortex
\begin{eqnarray}
W(C)=\mathrm{Tr}[UU...U]\rightarrow Tr[UU...\exp[{\frac{2\pi i n }{N}}]...U], 
\qquad n=1,2, \cdots, N-1.
\label{WU}
\end{eqnarray}
The place of insertion of $\exp({\frac{2\pi i n}{N}})$ is arbitrary, since it commutes with all elements of the gauge group.
If the minimal area is not pierced by any center vortex, a trivial center element, which is the unit matrix of the SU($N$) gauge group, is inserted. Thus, it does not contribute to the potential.

So far, only thin vortices were discussed. In fact, thin vortices give a correct linear potential at large distances. In general, the potential between two heavy sources is given by
\begin{equation}
V(R)=-\frac{A}{R} + KR +B,
\label{potential}
\end{equation}
where the coefficient of the linear term $K$ is called the string tension. The potential of thin vortices behaves according to the $N$-ality of the representation at large distances. The $N$-ality of a representation is the number of boxes in the Young Tableau of the corresponding representation mod N. For zero $N$-ality representations, the string tensions at large distances are zero and the representations with the same non-zero N-ality get the same string tensions.

The center vortex theory of quark confinement \cite{Thooft:1977} has received so many numerical supports \cite{numer:1997}. From this theory the lowest string tension for each N-ality is seen in the large distance limit due to screening. However it does not explain the intermediate string tensions especially for higher representations.
To understand the intermediate linear potentials of higher representations, which are observed in the lattice data, Faber \emph{et. al} modified the center vortex model to the thick center vortex model \cite{Faber:1997}. The linear potentials at intermediate distances have been obtained by this model. In this region, the string tensions of different group representations are roughly proportional to the Casimir ratios. Based on this modification, the center element $\exp({\frac{2\pi i n}{N}})$ in equation (\ref{WU}) is replaced by a group factor $G$ which is a unitary matrix and interpolates smoothly between the center elements. It would be equal to +I, if the core is outside the loop and $\exp({\frac{2\pi i n}{N}})$, if the core is contained entirely in $W(C)$. Thus
\begin{equation}
W(C)=\mathrm{Tr}[UU...U]\rightarrow \mathrm{Tr}[UU...G...U],
\end{equation}
where $G(x,s)=S\exp[i\vec{\alpha}^{n}_{C}(x)\cdot\vec{H}]S^{\dag}$. $\{H_{i},i=1,2,...,N-1\}$  are the generators spanning the Cartan
sub-algebra and $S$ is an element of the SU($N$) group representation $r$. The parameter $\alpha_{C}(x)$ describes the vortex flux distribution and depends on that fraction of the vortex core, which is enclosed by the loop $C$. Therefore, it depends on both the shape of the loop $C$, and the position x of the center of the vortex core, relative to the perimeter of $W(C)$, with the following conditions:

1. vortices which pierce the plane far outside the loop do not affect the loop.
In other words, for fixed $R$, as $x\rightarrow \infty$,
$\alpha \rightarrow 0$.

2. If the vortex core is entirely contained within the loop, then G would be equal to one of the center elements and $\alpha$ would get the maximum value.

3. As $R\rightarrow 0$ then $\alpha \rightarrow 0$.

Following the calculations of the reference \cite{Faber:1997}, the Wilson loop of the model of thick center vortices is obtained by
\begin{equation}
<W(C)> = \prod_{x} \{ 1 - \sum^{N-1}_{n=1} f_{n} (1 - \mathrm{Re}{g}_{r}
          [\vec{\alpha}^n_{C}(x)])\},
\label{WC}
\end{equation}
$x$ is the location of the center of the vortex and $n$ shows the vortex type. $f_{n}$ is the probability that any given unit is pierced by a vortex of type $n$. The number of vortices are equal to the number of non-trivial center elements of the gauge group and there are ($N-1$) vortices for the SU($N$) gauge group. However, vortices of type $n$ and $N-n$ are the same except that their fluxes flow in opposite directions. Thus, one may say that the number of vortices is int$(\frac{N}{2})$ for each SU($N$) gauge group, which is equal to the number of independent string tensions at large distances.
${g}_{r}$ is defined as
\begin{equation}
{g}_{r}[\vec{\alpha}] = \frac{1}{d_{r}} \mathrm{Tr} \exp[i\vec{\alpha} . \vec{H}],
\label{gr1}
\end{equation}
where $d_{r}$ is the dimension of the representation. 
Using the above average Wilson loop and equation (\ref{WV}), the potential between static sources is
 \begin{equation}
V(R) = -\sum_{x}\mathrm{ln}\{ 1 - \sum^{N-1}_{n=1} f_{n}
            (1 - \mathrm{Re}{g}_{r} [\vec{\alpha}^n_{C}(x)])\}.
\label{sigmac}
\end{equation}

This thick center vortex model could predict an appropriate large and intermediate distance behavior not only for the fundamental representation but also for the higher representations \cite{Faber:1997}, \cite{Deld:2000}, \cite{Deld:2005}. But, it has not predicted the Coulombic behavior. This is not strange, since only the non-trivial center elements have a significant role in the induced potential as confirmed by lattice calculations. We recall that the exponential in equation (\ref{W}) is caused by the non-trivial center elements. However, in the process of thickening the vortices, a flux distribution of the vortices has been chosen such that the slope of the profile causes also contributions near the trivial center element. But, most of the contributions to $\exp[-\sigma(C)A]$ still come from the non-trivial center elements, which leads to the asymptotic large distances behavior, and it gives the correct intermediate distance potentials, as well. We have been motivated to increase those contributions, which are close to the trivial center element to get the short distance potential.

In the next section, we introduce a linear profile function to describe the overlap of a vortex with a Wilson loop $W(C)$.

\section{Trivial and non-trivial center elements}

The group factor $G_r(\vec{\alpha}_C)$ in equation (\ref{WC}) parameterizes the influence of a vortex on a Wilson loop (\ref{WC}). We choose $n=1$ and $\vec{\alpha} . \vec{H}=\alpha_RH_{N-1}$, with $H_{N-1}=\mathrm{diag}(1,1,\cdots,N-1)/\sqrt{2N(N-1)}$. According to ref.~\cite{Faber:1997} the profile function $\alpha_R(x)$ is given  by
\begin{equation}
\alpha_{R}(x)= \pi\sqrt{\frac{2(N-1)}{N}}[1-\tanh(ay(x)+ \frac{b}{R})],
\label{alphar}
\end{equation}
 with
\begin{equation}
\label{yx}
 y(x)= \left \{ \begin{array}{ll}
x-R ~~ \mbox{for $|R-x| \leq |x|$ } \\
-x ~~ \mbox{for $|R-x| > |x|$ }
\end{array}
\right.
\end{equation}
where $R$ is the space-extent of a Wilson loop with infinite time-extent and $x$ the position of the vortex. $a$ and $b$  are free parameters of the model. $a$ is proportional to the inverse vortex core size.  The normalization factor in equation ~(\ref{alphar}) is chosen in such a way that for $R\to\infty$ and $x=R/2$ the first non-trivial center element of the gauge group SU($N$) is reached. In this case, the whole vortex is contained in the Wilson loop and ${g}_{r}[\vec{\alpha}]$ in equation (\ref{gr1}) reaches
\begin{equation}
{g}_{r}[\vec{\alpha}] = \frac{1}{d_{r}} \mathrm{Tr} \exp[\frac{2\pi i n}{N}].
\label{gr2}
\end{equation}

\begin{figure}[]
\begin{center}
\vspace{70pt}
\resizebox{0.47\textwidth}{!}{
\includegraphics{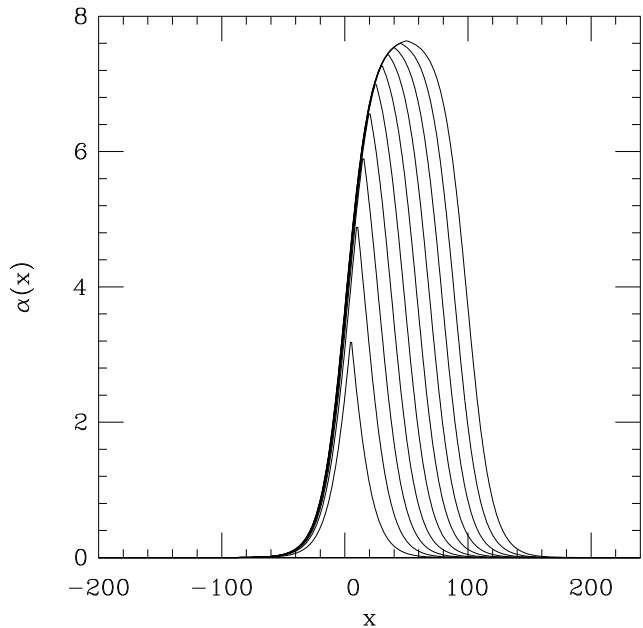}}
\vspace{-20pt}
\caption{\label{fluxall}
Profile function $\alpha_R(x)$ of Eq.~(\ref{alphar}) with $a=0.05$, $b=4$ and $R=10$ to $R=100$ in steps of $10$. If the vortex is far outside the Wilson loop, it does not influence the loop and if it is completely inside, $\alpha_{R}(x)$ reaches its maximum value $\sqrt{6}\pi=7.6953$. The latter case mainly happens for large $R$'s. }
\end{center}
\end{figure}

In Figure ~\ref{fluxall}, we study in SU(4) the vortex type one ($Z_1=i$), by plotting the vortex profile function $\alpha_R(x)$ of the fundamental representation for $R=10-100$ in steps of $10$ with the parameters $a=0.05$ and $b=4$. To produce this plot for each Wilson loop with the space-extent $R$, we put one leg of the Wilson loop on $x=0$ and the other leg on $x=R$. Then, $\alpha_R(x)$ in equation (\ref{alphar}) for each $x$, simply gives the contribution of the vortex   on the Wilson loop. Therefore, $\alpha_R(x)$ of each $x$ of this plot shows the effect of a vortex with a core at position $x$ on the phase of a Wilson loop with the space-extent $R$. If the vortex is far outside the Wilson loop, it does not influence the loop, if it is completely inside, $\alpha_{R}(x)$ reaches its maximum value $\sqrt{6}\pi=7.6953$. With this data set, the latter case mainly happens for larger $R$'s whereas for smaller $R$'s such as $R=10$ or $20$ the vortex is only partially inside the Wilson loop. The vortices that partially link the Wilson loop produce the intermediate linear potential and the short distance potential.  The asymptotic large distance behavior is observed for the case when the vortex is completely inside the loop.

\begin{figure}[]
\begin{center}
\vspace{70pt}
\resizebox{0.47\textwidth}{!}{
\includegraphics{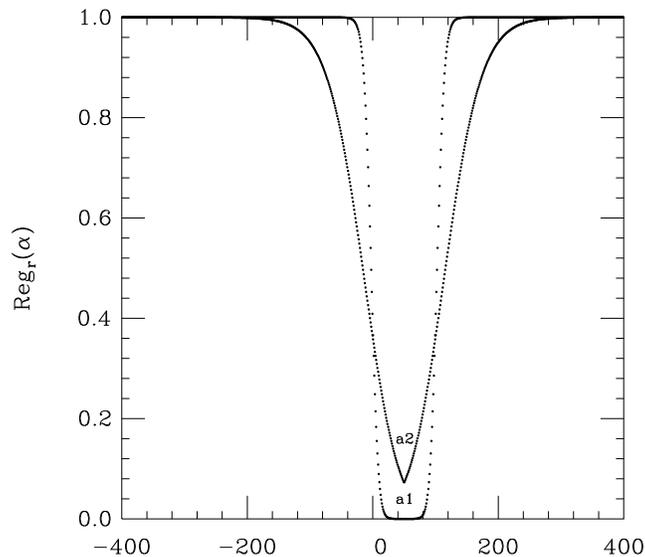}}
\vspace{-20pt}
\caption{\label{gralpha1}
$\mathrm{Re}{g}_{r}[\vec{\alpha}]$ of vortex type one is plotted versus $x$ for the fundamental representation of the gauge group SU(4) for a Wilson loop with space extent $R=100$ and the data set $a_1:(a=0.05,b=4)$ and $a_2:(a=0.01,b=1)$. $\mathrm{Re}{g}_{r}[\vec{\alpha}]$ is equal to $1$, when no vortex links the loop and it is $0$, when the vortex is completely inside the loop.}
\end{center}
\end{figure}

\begin{figure}[]
\begin{center}
\vspace{70pt}
\resizebox{0.47\textwidth}{!}{
\includegraphics{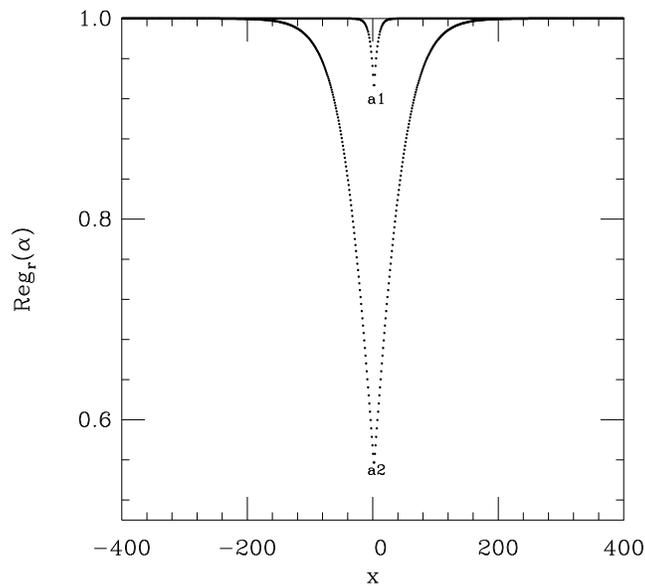}}
\vspace{-20pt}
\caption{\label{gralpha-both}
$\mathrm{Re}{g}_{r}[\vec{\alpha}]$ of vortex type one is plotted versus $x$ for the fundamental representation of the gauge group SU(4) for a Wilson loop with space extent $R=5$. The plot labeled by $a_{1}$ refers to the parameters $a=0.05$, $b=4$ and the one labeled by $a_{2}$ to $a=0.01$, $b=1$. The distribution near the trivial center element is increased for the second case. In contrast to figure \ref{gralpha1} $\mathrm{Re}{g}_{r}[\vec{\alpha}]$ does not reach zero, since $R$ is smaller than the vortex cores size and the vortex never links the loop completely.
}
\end{center}
\end{figure}

We have studied $Re{g}_{r}[\vec{\alpha}]$ for different $a$'s for a couple of different distances, $R$, and discuss the effect of the vortex core size on $Re{g}_{r}[\vec{\alpha}]$ and on the contributions of trivial and non-trivial center elements. Figure \ref{gralpha1} represent $\mathrm{Re}{g}_{r}[\alpha_R]$ of the vortex type one of the SU($4$) gauge group, for $R=100$ for the parameter sets $a_1:(a=0.05,b=4)$ and $a_2:(a=0.01,b=1)$. If the vortex core is far outside the Wilson loop, then $\mathrm{Re}{g}_{r}[\vec{\alpha}]=1$. For the data set $a_1$, the vortex core is completely inside the loop around $x\approx\frac{R}{2}=50$ and contributes with $\mathrm{Re}{g}_r[\alpha_R]=0$ , but for the data set $a_2$ where we have used a bigger core for the vortex, the vortex is not completely inside the loop even for $R=100$. In general, $Re{g}_{r}[\vec{\alpha}]$ changes between $1$, when the vortex does not link the loop, and zero, when the vortex is completely inside the loop. If we plotted $Re{g}_{r}[\vec{\alpha}]$
for the thin vortex model, we would get only $1$ and $0.0$. Thus, increasing the thickness of the vortex, implies a distribution between $1$ and $0$.  Figure \ref{gralpha-both} shows $Re{g}_{r}[\vec{\alpha}]$ for $R=5$ for the two parameter sets $a_1:(a=0.05,b=4)$ and $a_2:(a=0.01,b=1)$. As in figure \ref{gralpha1} if the vortex core is far outside the Wilson loop, then $\mathrm{Re}{g}_{r}[\vec{\alpha}]=1$. It is observed that for $R=5$ for both data sets, the vortex is only  partially inside the Wilson loop. 

From figure \ref{gralpha1}, it is observed that for the case when we use the data set $a_1$, we have a very fast transition between $1$ and $0$ for $\mathrm{Re}{g}_{r}[\vec{\alpha}]$. On the other hand, for the data set $a_2$ the transition happens very slowly such that the number of points between $0$ and $1$ is increased significantly. For figure \ref{gralpha-both} where $R=5$ and $\mathrm{Re}{g}_{r}[\vec{\alpha}]$ does not reach to zero, the change of the distribution is also observed, especially when 
$\mathrm{Re}{g}_{r}[\vec{\alpha}]$ is close to $1$ where $1$  corresponds to the trivial center element. Recall that we wish to obtain a Coulombic behavior for the short distance region, $R<10$ in the given example. The idea is to add back $W_0$ to the Wilson loop in equation (\ref{W}). 
As mentioned in section $2$, the exponential part of equation (\ref{W}) leads to a linear potential and from the center vortices model, only the non-trivial center elements are needed to reproduce the linear part. We have been motivated to reproduce the Coulombic part by increasing the role of the trivial center element. It is obvious that using the trivial center element like a thin vortex would mean that one inserts an $I$ some where between $U$'s in equation $W(C)=\mathrm{Tr}[UU...U]$ which leads to $\mathrm{Re}{g}_{r}[\vec{\alpha}]=1$. But since we are using thick center vortices, inserting a trivial center element or a ``trivial vortex" may be considered like the real vortices and therefore those contributions which are close to $1$ may influence the potential. It is like the case where thickening the vortices let us have a linear potential for the intermediate distances. 

Back to Figures \ref{gralpha1} and \ref{gralpha-both}, we expect to see a different behavior for the short distances with the new distribution produced from the data set $a_2$.

In the next section, we calculate the potential between static sources with the new data set $a_2$ and show that using the contributions close to the trivial center element or the ``trivial vortex" leads to a Coulombic potential.

\section{Restoring short distance potentials}

\begin{figure}%f1
\begin{center}
\vspace{70pt}
\resizebox{0.47\textwidth}{!}{
\includegraphics{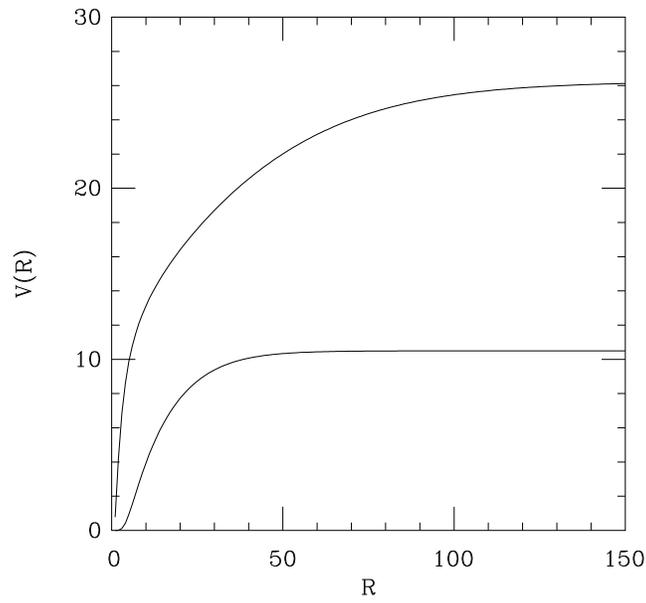}}
\vspace{-20pt}
\caption{\label{coul-pot}
Potentials between two heavy adjoint sources of the SU(3) gauge group with the parameters $a=0.05$, $b=4$ and $a=0.02$, $b=1$. For the second data set corresponding to the upper curve a Coulombic behavior appeared at small distances.
}
\end{center}
\end{figure}

Figure \ref{coul-pot} shows the potentials between two heavy adjoint sources of the SU(3) gauge group with the parameters $a=0.05$, $b=4$ and $a=0.02$, $b=1$. As one can see in the figure, a Coulombic behavior appears at small distances for the second data set.
Therefore, with data set $a_{2}$, we see clear evidence for Coulombic behavior at short distances, due to the enhancement of the vortex effect especially the trivial center element in this regime. 

\begin{figure}%f1
\begin{center}
\vspace{70pt}
\resizebox{0.47\textwidth}{!}{
\includegraphics{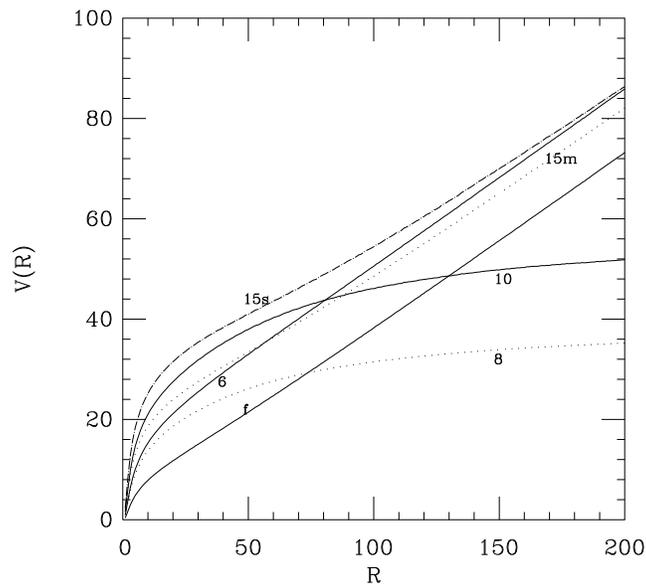}}
\vspace{-20pt}
\caption{\label{SU3}
Potentials between heavy sources of different SU(3) representations. Potentials at short distance show a Coulombic behavior. Both, the intermediate and short distance potentials are proportional to the Casimir scaling.
}
\end{center}
\end{figure}

\begin{figure}%f1
\begin{center}
\vspace{70pt}
\resizebox{0.47\textwidth}{!}{
\includegraphics{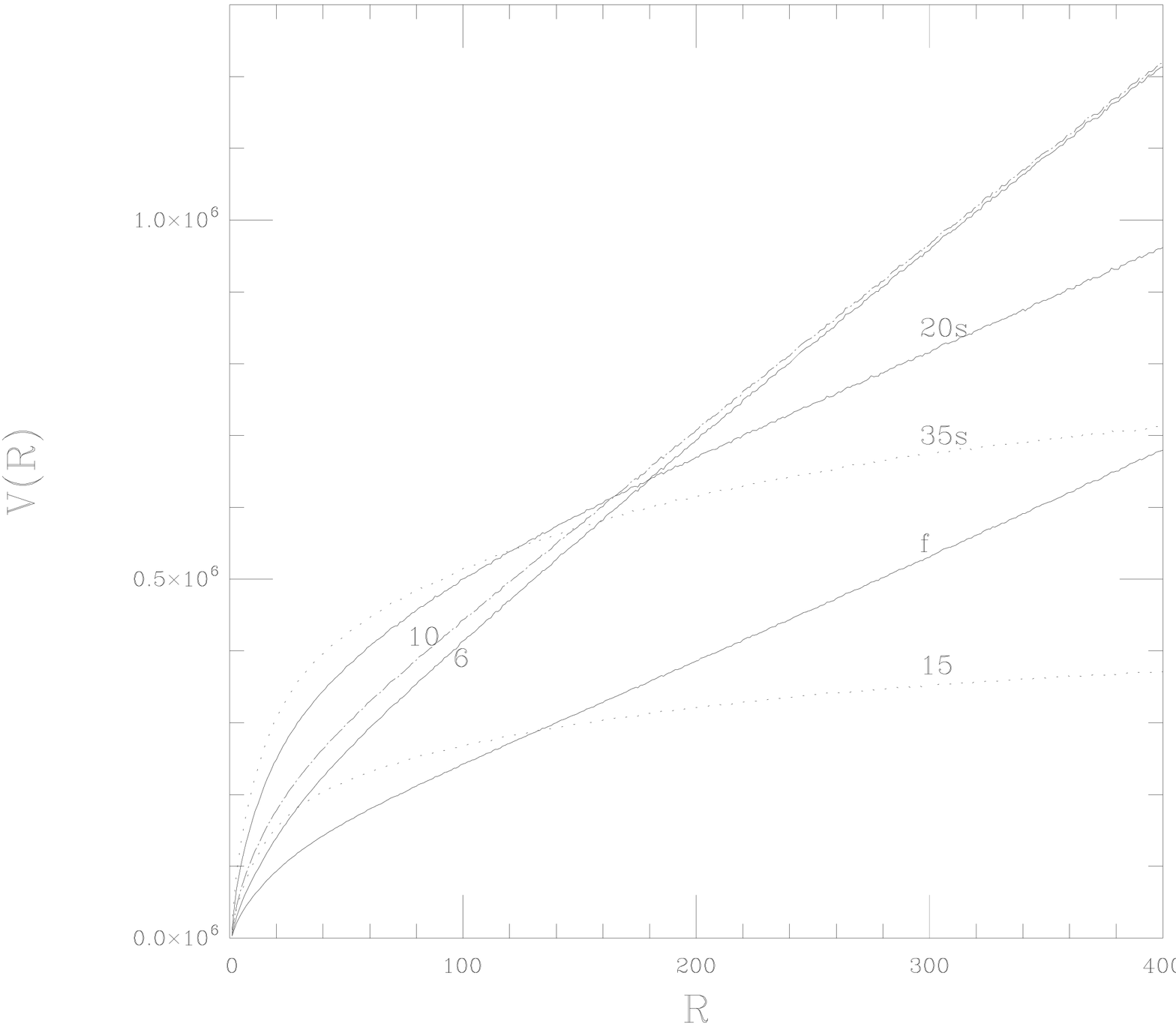}}
\vspace{-20pt}
\caption{\label{SU4}
The same as figure \ref{SU3}, but for SU(4) gauge group.
}
\end{center}
\end{figure}

It is seen that by increasing the role of the trivial center element in the model of thick center vortices, $W_{0}$ of equation (\ref{W}) is enhanced in the induced potential of the model.  However, the linear potential at intermediate distances has lost its previous shape and some concavity is observed in that area. To overcome this problem and the fact that the length of the intermediate linear part has been decreased, we let the vortex core size fluctuate by varying the parameter $a$ with a Gaussian distribution around an appropriate central value which gives the Coulombic potential. The tails of the Gaussian have been removed to avoid core sizes far from this central value. In order to save the general form of the potentials, the parameter $b$ has been scaled with $a$. The idea of using a distribution for the parameter $a$ to remove the concavity of the potential has been experienced in one of our earlier works \cite{Deldar:2007}. The satisfactory results of that study has encouraged us to fluctuate the parameter $a$ which is proportional to the inverse of he vortex core size. However, we have been very careful to not change the shape of the flux which leads to the Coulombic behavior. This has been done by choosing a Gaussian distribution for the parameter $a$ without the tails to avoid the core sizes which are far from the original values of the data set $a_{2}$. With this trick, the concavity can be decreased and a reasonable behavior is achieved for intermediate and short distances regime. Several values for $a$ have been tried out and finally, we have chosen $a=0.01$ and $b=1$  for the vortex type one of the SU(4) gauge group and $a=0.005$ and $b=10$ for the vortex type two.  For SU(3), $a=0.02$ and $b=1$ gives the best results. $w$, the width of the Gaussian has been chosen such that $w=\frac{a}{2}$. We should mention that a vast variety of free parameters may lead to Coulombic type potential, but the point is that we want to get Casimir scaling for both short and intermediate distances. The appropriate parameters may suggest some real profiles for the vortices in agreement with lattice data. Figures~\ref{SU3} and \ref{SU4} display the potentials for different representations including the fundamental representation for the SU(3) and SU(4) gauge groups. 

Tables 1 and 2 show the results of fitting the model to equation ~(\ref{potential}), for the gauge group SU(3) and SU(4), respectively. Qualitative agreement with Casimir scaling is obtained for all the coefficients of the equation.
The fact that we could fit the potentials with equation ~(\ref{potential}), $V(R)=-\frac{A}{R} + KR +B$, and observing a rough agreement with Casimir scaling could be an evidence of obtaining the Coulombic behavior for the short distance. Another interesting point is the approximate agreement of $B$ with Casimir scaling. This term represents a perimeter term which its agreement with Casimir scaling automatically happens if the Casimir scaling holds for the Coulomb and linear terms. Finally we would like to mention that although the results are only roughly consistent with Casimir scaling, the agreement is not worse than what we have found  by using this model for SU($3$) and SU($4$) gauge groups potentials \cite{Deld:2000}, \cite{Deld:2005}, \cite{Rafi:2007} or even the original work obtained from the model for SU($2$) potentials \cite{Faber:1997}.  In fact, so far the results have been qualitatively in agreement with Casimir scaling. However the main features of the potentials between static quarks including a Coulombic behavior for short distances, linear potential at intermediate distances and correct large distance behavior based on the N-ality of the representations are obtained successfully by the thick center vortices model.

\section{Conclusions}

We have tried to extend the thick center vortex model to find the potentials between heavy sources not only for intermediate and large distances, but also for short distances, where a Coulombic behavior is expected. We have increased the role of vortex fluxes piercing Wilson loops with contributions, which are close to the trivial center element. With this method, we obtain an appropriate potential for the short and intermediate distances. To obtain a convex potential and an acceptable size for the intermediate regime, the vortex core size has been fluctuated by using a Gaussian distribution. The Coulombic behavior can be obtained for several values of the free parameters of the chosen profile. However, the proportionality with Casimir scaling is obtained by carefully chosen vortex core sizes and by fluctuations of them very smoothly.

The requirement that the results agree with lattice calculations may lead to valuable information about the vortex profile, including the distribution and the vortex core size.

\begin{table}%t2
\tabcolsep=2pt
%\begin{ruledtabular}
\caption[]{\label{table2}
The fits of equation ~(\ref{potential}) to the potentials for the gauge group SU(3) are shown. $A_{r}/A_{f}$ is the ratio of the Coulombic coefficient of the representation $r$ to the fundamental representation, $K_{r}/K_{f}$ is the ratio of the string tensions, $B_{r}/B_{f}$ the ratio of the constant terms and $C_{r}/C_{f}$ is the Casimir ratio. Qualitative agreement of the coefficients with  Casimir scaling  is
observed. The errors of the fits are shown in parentheses.}
\begin{center}
\begin{tabular}{@{}llcccccc@{}}
\hline
Repn.&    $8$ & $6$ & $15a$&  $10$ &  $15s$
\\
\hline
$(n,m)$    & $(1,1)$ & $(2,0)$ &  $(2,1)$ & $(3,0)$ & $(4,0)$ \\
$C_{r}/C_{f}$   &  $2.25$ &  $2.5$ &  $4$ &   $4.5$ &  $7$ \\
$A_{r}/A_{f}$   &  $2.09(21)$ &   $2.31(23)$ & $2.85(27)$  &  $3.37(32)$ & $4.04(37)$\\
$K_{r}/K_{f}$   &  $1.31(8)$ &   $1.49(9)$ & $1.52(11)$  &  $1.62(10)$ &  $1.63(12)$  \\
$B_{r}/B_{f}$   &  $2.24(12)$ &   $2.45(12)$ & $3.23(15)$  &  $3.84(17)$ &  $4.93(20)$  \\
\hline
\end{tabular}
%\end{ruledtabular}
\end{center}
\end{table}

\begin{table}%t2
\tabcolsep=2pt
%\begin{ruledtabular}
\caption[]{\label{table1}The same as table 1, but for the gauge group SU(4).}
\begin{center}
\begin{tabular}{@{}llcccccc@{}}
\hline
Repn.&    $6$ & $15$ & $10$&  $20_{s}$ &  $35_{s}$
\\
\hline
$(n,m)$    & $(2,0)$ & $(1,1)$ &  $(2,0)$ & $(3,0)$ & $(4,0)$ \\
$C_{r}/C_{f}$   &  $1.33$ &  $2.13$ &  $2.4$ &   $4.2$ &  $6.4$ \\
$A_{r}/A_{f}$   &  $1.38(2)$ &   $2.05(4)$ & $2.26(4)$  &  $3.42(7)$ & $4.59(11)$\\
$K_{r}/K_{f}$   &  $1.51(1)$ &   $1.56(2)$ & $1.76(2)$  &  $2.38(3)$ &  $2.79(4)$  \\
$B_{r}/B_{f}$   &  $1.34(1)$ &   $2.18(3)$ & $2.37(3)$  &  $3.77(6)$ &  $5.22(9)$  \\
\hline
\end{tabular}
%\end{ruledtabular}
\end{center}
\end{table}

\section{Acknowledgments}

We would like to thank Manfried Faber and Denise Neudecker for the very helpful discussions. This work is  supported by the research council of the University of Tehran.

\end{document}